# Simple derivation of the first cumulant for the Rouse chain

V. Lisy [a], B. Brutovsky [b], J. Tothova [b]

[a] Department of Physics, Technical University of Kosice,
Park Komenskeho 2, 042 00 Kosice, Slovakia

[b] Institute of Physics, P.J. Safarik University, Jesenna 5, 041 54 Kosice, Slovakia

**Abstract.** A simple analytic expression for the first cumulant of the dynamic structure factor of a polymer coil in the Rouse model is derived. The obtained formula is exact within the usual assumption of the continuum distribution of beads along the chain. It reflects the contributions to the scattering of light or neutrons from both the internal motion of the polymer and its diffusion, and is valid in the whole region of the wave-vector change at the scattering.

__________________________________________________________________________________

The Rouse model [1] is a fundamental theory for the dynamics of a polymer chain in liquids. Despite its simplicity, it finds a number of applications, e.g., in the description of polymer dynamics in situations when the hydrodynamic interactions between the monomers are screened out by surrounding polymers [2]. Such effects are extensively studied both theoretically and in experiments, first of all by quasielastic scattering of light and neutrons. The central quantities of interest in these studies are the dynamic structure factor (DSF) and its first cumulant $\Gamma$, approximate formulas for which have been obtained already in the classic de Gennes' work forty years ago [3]. Later, Akcasu *et al.* [4] derived closed expressions for the first cumulant both in the free-draining limit and with hydrodynamics included, but these formulas are too complicated to be used in the interpretation of experimental data. As discussed in the recent paper [5], they are given as summations over the number of beads in the chain and produce simple results only in the limit of large scattering vectors. Due to this, experimenters use limiting results (the well-known $q^2$ and $q^4$ laws [2]) that are correct at small and large scattering vectors $q$, but not valid in the intermediate region of $q$. In the mentioned work [5], an analytic expression for a different quantity, called the mean decay rate of the internal autocorrelation function of a Gaussian coil, has been obtained in the free-drainage limit for chains of length much larger than the size of the monomer. As will be shown below, this quantity (if also the diffusion of the whole coil is taken into account) can be related to $\Gamma$. To our knowledge, however, the exact analytic formula for the Rouse cumulant itself has not been explicitly given in the literature.

In this note, the first Rouse cumulant to the DSF is exactly calculated. The found expression as well as the method of its obtaining are very simple. The cumulant, reflecting the contributions to the scattering from both the internal motion within the coil and its diffusion as a whole, describes the experimentally measured initial slope of the DSF. For a long Gaussian polymer coil mapped by $N$ beads, the DSF is determined as [2]

$$G(q,t) = \frac{1}{N} \sum_{n,m} \exp[-q^2 \Phi_{nm}(t)], \tag{1}$$

$$\Phi_{nm}(t) = Dt + 2\sum_{p=1}^{\infty}\left[\psi_p(0)\left(\cos^2\frac{\pi np}{N} + \cos^2\frac{\pi mp}{N}\right) - 2\psi_p(t)\cos\frac{\pi np}{N}\cos\frac{\pi mp}{N}\right]. \tag{2}$$

Here, $D = k_B T/(3\pi N\eta d)$ is the diffusion coefficient of the coil ($\eta$ is the solvent viscosity and $d$ the diameter of the bead), the indexes $m$ and $n$ change from 0 to $N$, and $\psi_p(t)$ is the time correlation function of the Fourier components of the bead radius vector [2]. As usually, the continuum approximation for the distribution of beads is used. This implies that the number of the internal modes in Eq. (2) is infinite. In the Rouse case, $\psi_p(t) = \psi_p(0)\exp(-t/\tau_p)$ with $\psi_p(0) = Nb^2/(6\pi^2 p^2)$ and the relaxation times $\tau_p = N^2 b^2 d\eta/(\pi b k_B T p^2)$, where $b$ is the mean-square distance between the beads along the chain [2]. The static structure factor is [2]

$$G(q,0) = 2N\kappa^{-1}\left\{1 - \kappa^{-1}\left[1 - \exp(-\kappa)\right]\right\}, \qquad \kappa \equiv (qR_G)^2. \tag{3}$$

Here, $R_G = (Nb^2/6)^{1/2}$ is the radius of gyration of the coil. The first cumulant of the DSF is defined as [2]

$$\Gamma = -G(q,0)^{-1}[\partial G(q,t)/\partial t]_{t=0}. \tag{4}$$

At $t = 0$, $\Phi_{nm}(0) = b^2|n-m|/6$ and the derivative $d\Phi_{nm}(t)/dt$ can be expressed in the form

$$\left(\frac{d\Phi_{nm}(t)}{dt}\right)_{t=0} = D\left[1 + 2\sum_{p=1}^{\infty}\cos\frac{\pi np}{N}\cos\frac{\pi mp}{N}\right], \tag{5}$$

where we have used $\psi_p(0)/\tau_p = D/2$. The time derivative of the DSF is

$$\left(-\frac{\partial G}{\partial t}\right)_{t=0} = \frac{q^2}{N}\sum_{nm}\left(\frac{d\Phi_{nm}(t)}{dt}\right)_{t=0}\exp(-q^2 R_G^2 |n-m|/N). \tag{6}$$

Calculating the sum, we convert it to the double integral. The integral is easily found noting that the quantity in the square brackets in Eq. (5) is proportional to the delta function,

$$N\delta(n-m) = 1 + 2\sum_{p=1}^{\infty}\cos\frac{\pi np}{N}\cos\frac{\pi mp}{N}, \tag{7}$$

represented in the interval [0, $N$] through the set of the orthonormal functions $\psi_p(n) = (2-\delta_{p0})^{1/2} N^{-1/2}\cos(\pi np/N)$, $p = 0, 1, 2,...$ Thus, one finds

$$-\left.\frac{\partial G}{\partial t}\right|_{t=0} = k^2 D \int_0^N dn \int_0^N dm\, \delta(n-m) \exp\left(-\frac{q^2 R_G^2}{N}|n-m|\right) = q^2 ND. \tag{8}$$

Then the first cumulant $\Gamma$ from Eqs. (4), (3), and (8), normalized to $q^2 D$, is (Fig. 1)

$$\gamma_R = \frac{\Gamma}{q^2 D} = \frac{1}{2} \frac{\kappa^2}{\kappa + \exp(-\kappa) - 1}, \quad \kappa = (qR_G)^2, \tag{9}$$

with the well-known limiting cases $\gamma_R \approx 1 + \kappa/3 + \ldots$ ($\kappa \to 0$) and $\gamma_R \approx \kappa/2$ ($\kappa \gg 1$) [2].

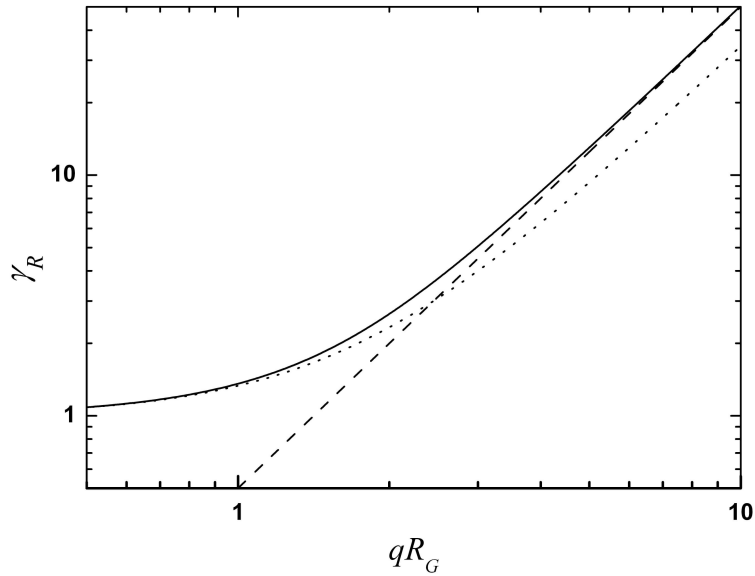

**Fig. 1.** The normalized first cumulant (9) (full line) and its high- and low-$q$ limits for a Rouse polymer (dashed and dotted lines, respectively). The analytic expression (9) is indistinguishable from the numerical calculation from Eqs. (4), (1), and (2).

In Ref. [5], the mean decay rate of the internal autocorrelation function of a Gaussian coil was calculated using the definition

$$K_1 = -\left.\frac{\partial [\ln(G_{\text{int}} - G_{\text{int},\infty})]}{\partial t}\right|_{t=0} = -\frac{G'_{\text{int},0}}{G_{\text{int},0} - G_{\text{int},\infty}}, \tag{10}$$

where $G_{\text{int}}(q,t)$ is given by Eq. (1) at $D = 0$, $G'_{\text{int},0}$ is the derivative of $G_{\text{int}}(q,t)$ evaluated at $t = 0$, $G_{\text{int},0}$ coincides with $G(q,0)$, and $G_{\text{int},\infty} = \lim_{t \to \infty} G_{\text{int}}(q,t)$:

$$G_{\text{int},\infty} = \frac{1}{N} \sum_{nm} \exp\left\{-\frac{Nb^2 q^2}{3\pi^2} \sum_{p=1}^{\infty} \frac{1}{p^2} \left(\cos^2 \frac{\pi np}{N} + \cos^2 \frac{\pi mp}{N}\right)\right\}. \tag{11}$$

Hence, with the help of the sum $\sum_{p=1}^{\infty} p^{-2} \cos^2(pz) = (z^2 - z\pi + \pi^2/3)/2$ [6], one finds

$$G_{int,\infty} = \frac{1}{N} \exp\left(-\frac{2}{3}\kappa\right) \left(\sum_n \exp\left[\frac{\kappa n}{N}\left(1-\frac{n}{N}\right)\right]\right)^2 = \frac{\pi N}{\kappa} \exp\left(-\frac{\kappa}{6}\right) \left[\text{erf}\left(\frac{\sqrt{\kappa}}{2}\right)\right]^2, \qquad (12)$$

where the sum was again replaced with integral ($\kappa$ is from Eq. 9). Since $\Gamma$ is related to $K_1$ by the equation $\Gamma = q^2 D + K_1(1 - G_{int,\infty}/G_{int,0})$ (obtained after the derivation of $G(q,t)$ and using the definition (10) of $K_1$), one easily recovers $K_1$ from Ref. [5] without complicated calculations of $G'_{int,0}$. However, the quantity that is actually measured in experiments is not $K_1$ but the cumulant (4) of the total DSF, given by a much simpler formula (9). Note that if $q^2 b^2/6 \ll 1$, $\Gamma$ from Eq. (9) can be also identified with the half-width of the quasi-elastic peak in the scattering law $G(q,\omega)$ found in the paper by Akcasu and Gurol [7].

Finally, it should be stressed that the results presented here and in Ref. [5] are valid if in the equations of motion of the chain the continuum approximation for the position vectors of beads is used. In building discrete variants of the theory, with a finite number of internal modes, the relaxation times and eigenvectors different from the usual textbook results [2] should be used [8]. As shown in Ref. [8], these results apply if the mode number $p$ is much smaller than the number of beads $N$. Nevertheless, the continuum approximation is well substantiated for the long-time polymer dynamics [2], in which only case the bead-spring models are reasonable.

Acknowledgment. We are indebted to T. Matsoukas for illuminating remarks on his work [5]. VL and BB acknowledge the support from the Scientific Grant Agency of the Slovak Republic (VEGA).